\magnification=1100
\baselineskip=15truept
\hsize=6.25truein
\vsize=9truein
\emergencystretch=2truepc\raggedbottom\frenchspacing\parskip=0pt
\font\Tfont=cmbx12
\def\T#1\par{\centerline{\Tfont#1}}
\newcount\secno\newcount\intro\newcount\vaidboxes\newcount\pcer
\newcount\opretro\newcount\qmcc\newcount\uinex
\newcount\mapicos\newcount\conclu
\def\section#1#2\par{\advance\secno1#1=\secno\bigskip\noindent%
    {\bf\number\secno.\ #2}\par\smallskip}
\def\longsection#1#2\par#3\par{\advance\secno1#1=\secno%
    \bigskip{\bf\noindent\number\secno.\ #2\par
    \noindent#3\par}\smallskip}
\def\affil#1\par{\vskip11truept\centerline{\it#1}}
\font\nnfont=cmr7\newcount\nnref
\long\def\fnote#1{\global\advance\nnref1%
    \footnote{\raise1ex\hbox{\nnfont\number\nnref}}{\vskip-13truept#1\vskip-13truept}}
\def\absosq#1{\bigl\vert#1\bigr\vert^2} 
\def\ket#1{\vert#1\rangle}

\def\ketbra#1#2{\vert#1\rangle\langle#2\vert}
\def\sandwich#1#2#3{\langle#1\vert#2\vert#3\rangle}

\T Quantum Mechanics and Elements of Reality

\vskip22truept\centerline{\rm Ulrich 
Mohrhoff\hskip2pt\footnote{\raise1ex\hbox{\nnfont a)}}{E-mail: 
ujm@auroville.org.in}}

\affil Sri Aurobindo Ashram, Pondicherry 605002, India

\vskip30truept\centerline{\bf Abstract}
{\leftskip=\parindent\rightskip=\parindent\noindent It is widely accepted that a 
Born probability of~1 is sufficient for the existence of a corresponding element 
of reality. Recently Vaidman has extended this idea to the ABL probabilities of 
the time-symmetrized version of quantum mechanics originated by Aharonov, 
Bergmann, and Lebowitz. Several authors have objected to Vaidman's
time-symmetrized elements of reality without casting doubt on the widely 
accepted sufficiency condition for `ordinary' elements of reality. In this paper I 
show that while the proper truth condition for a quantum counterfactual is an 
ABL probability of~1, neither a Born probability of~1 nor an ABL probability 
of~1 is sufficient for the existence of an element of reality. The reason this is so 
is that the contingent properties of quantum-mechanical systems are extrinsic. 
To obtain this result, I need to discuss objective probabilities, retroactive 
causality, and the objectivity or otherwise of the psychological arrow of time. 
One consequence of the extrinsic nature of quantum-mechanical properties is 
that quantum mechanics presupposes property-defining actual events (or states 
of affairs) and therefore cannot be called upon to account for their occurrence 
(existence). Neither these events nor the correlations between them are 
capable of explanation, the former because they are causal primaries, the latter 
because they are fundamental: there are no underlying causal processes. 
Causal connections are something we project onto the statistical correlations, 
and this works only to the extent that statistical variations can be ignored. 
There are nevertheless important conclusions to be drawn from the
quantum-mechanical correlations, such as the spatial nonseparability of the 
world.\par}

\vskip20truept\section{\intro}Introduction

Recently the concept of time-symmetric elements of reality, introduced by 
Vaidman (1996a, 1997), stirred up a lively controversy which culminated in the 
joint publication of two papers in this journal (Kastner, 1999; Vaidman 1999). 
Using the standard formalism of standard quantum mechanics, one calculates 
the Born probability
$$
P_B(a_i)=|\sandwich{\Psi}{{\bf P}_{A=a_i}}{\Psi}|,\eqno{(1)}
$$
where the operator ${\bf P}_{A=a_i}$ projects on the subspace corresponding 
to the eigenvalue $a_i$ of the observable $A$. $P_B(a_i)$ is generally 
regarded as {\it the} probability with which a measurement of $A$ performed 
after the `preparation' of a system $S$ in the `state' $\ket{\Psi}$ yields the 
result $a_i$. But $P_B(a_i)$ is not the only such probability. Using a
nonstandard formulation of standard quantum theory called
time-symmetrized quantum theory (Aharonov and Vaidman, 1991; Vaidman, 
1998), one calculates the ABL probability
$$
P_{ABL}(a_i)={\absosq{
\sandwich{\Psi_2}{{\bf P}_{A=a_i}}{\Psi_1}
}\over
\Sigma_j\absosq{
\sandwich{\Psi_2}{{\bf P}_{A=a_j}}{\Psi_1}
}}.\eqno{(2)}
$$
ABL probabilities were first introduced in a seminal paper by Aharonov, 
Bergmann, and Lebowitz (1964). In this paper it was shown that $P_B(a_i)$ 
can also be thought of as the probability with which a measurement of $A$, 
performed {\it before} what may be called the `retroparation' of $S$ in the 
`state' $\ket{\Psi}$, yields the result $a_i$. Further it was shown that if a system 
is `prepared' at the time $t_1$ and `retropared' at the time $t_2$ in the 
respective `states' $\ket{\Psi_1}$ and $\ket{\Psi_2}$, the probability with which 
a measurement of $A$ performed at an intermediate time $t_m$ yields (or 
would have yielded) the result $a_i$, is given by $P_{ABL}(a_i)$.\fnote{%
The $\Psi$'s in (2) are related to the `pre-/retropared' $\Psi$'s via unitary 
transformations\break $U(t_m-t_1)$ and $U(t_m-t_2)$.}

Born probabilities can be measured (as relative frequencies) using preselected 
ensembles (that is, ensembles of identically `prepared' systems). ABL 
probabilities can be measured using pre- and postselected ensembles (that is, 
ensembles of systems that are both identically `prepared' and identically 
`retropared'). If the Born probability $P_B(a_i,t)$ of obtaining the result $a_i$ at 
time $t$ is equal to~1, one feels justified in regarding the value $a_i$ of the 
observable $A$ as a property that is actually possessed at the time $t$, that is, 
one feels justified in assuming that at the time $t$ there is is an {\it element of 
reality} corresponding to the value $a_i$ of the observable $A$ irrespective of 
whether $A$ is actually measured. Redhead (1987) has expressed this feeling 
as the following `sufficiency condition':

\medskip{\leftskip=\parindent\rightskip=\parindent\noindent
(ER1) If we can predict with certainty, or at any rate with [Born] probability one, 
the result of measuring a physical quantity at time $t$, then at the time $t$ 
there exists an element of reality corresponding to the physical quantity and 
having a value equal to the predicted measurement result.\par}\medskip

The controversy about time-symmetric elements of reality arose because it 
appeared that Vaidman (1993) made the same claim with regard to ABL 
probabilities:

\medskip{\leftskip=\parindent\rightskip=\parindent\noindent
(ER2) If we can infer with certainty [that is, with ABL probability one] that the 
result of measuring at time $t$ an observable $A$ is $a$, then at the time $t$ 
there exists an element of reality $A=a$.\par}\medskip

In response to criticism by Kastner and others (Kastner, 1999; and references 
therein), Vaidman (1999) clarified that he intended the term `element of reality' 
in a `technical' rather than `ontological' sense: saying that there is an element 
of reality $A=a$ is the same as saying that if $A$ is measured, the result is 
certain to be $a$. In other words, (ER2) is a tautology: if $A=a$ is certain to be 
found then $A=a$ is certain to be found. In formulating (ER2), Vaidman does 
not affirm the existence of an element of reality {\it irrespective} of whether $A$ 
is actually measured. (ER2) {\it defines} what it means to affirm the existence 
of an element of reality corresponding to $A=a$. To say that there is such an 
element of reality is to affirm the truth of a conditional, not the existence of an 
actual situation or state of affairs. Since ordinarily the locution `element of 
reality' refers to an actual state of affairs, Vaidman's terminological choice was 
unfortunate and has mislead many readers. But beyond that, his reading of 
(ER2) is unobjectionable. I shall, however, stick to the ordinary, ontological 
meaning of `element of reality.' In what follows, (ER2) is to be understood 
accordingly, that is, as affirming an actual state of affairs (the existence of an 
`ordinary' element of reality) just in case the corresponding ABL probability is 
one. Hence my showing that (ER2), thus understood, is false, has no bearing 
on Vaidman's reading of (ER2). A definition cannot be false.

The aim of this paper is to show that not only (ER2) but also (ER1) is false. In 
Sec.~2 I discuss the three-box {\it gedanken} experiment due to Vaidman 
(1996b). By calculating the ABL probabilities associated with different versions 
of this experiment I show that positions are extrinsic, and that, consequently, 
(ER1) and (ER2) are both false. Since the validity of arguments based on
time-symmetric quantum counterfactuals is open to debate, in Sec.~3 I show 
without making use of ABL probabilities that positions are extrinsic and that 
(ER1) is false. This leads to the conclusion that the measurement problem is a 
pseudoproblem, and that all that ever gets objectively entangled is 
counterfactuals. In Sec.~4 I establish the cogency of the argument of Sec.~2 
by showing that the proper condition for the truth of a quantum counterfactual is 
$P_{ABL}=1$. This necessitates a discussion of objective probabilities, 
retroactive causality, and the objectivity or otherwise of the psychological arrow 
of time. In Sec.~5 I argue that since quantum mechanics presupposes the 
occurrence/existence of actual events and/or states of affairs, it cannot be 
called upon to account for the emergence of `classicality.' What is more, if 
quantum mechanics is as fundamental as its mathematical simplicity and 
empirical success suggest, the property-defining events or states of affairs 
presupposed by quantum mechanics are causal primaries -- nothing accounts 
for their occurrence or existence.

If this is correct, the remaining interpretative task consists not in explaining the 
quantum-mechanical correlations and/or correlata but in understanding what 
they are trying to tell us about the world. I confine myself to pointing out, in 
Sec.~6, the most notable implications of the diachronic correlations, viz., the 
existence of entities of limited transtemporal identity, objective indefiniteness, 
and the spatial nonseparability of the world. The extrinsic nature of positions, 
finally, appears to involve a twofold vicious regress. Its resolution involves 
macroscopic objects, which are defined and discussed in Sec.~7. Section~8 
concludes with a remark on the tension of contrast between objective 
indefiniteness and the inherent definiteness of language.

\section{\vaidboxes}The Lesson of the Three-Box Experiment

In the following I present a somewhat different but conceptually equivalent 
version of Vaidman's (1996b) three-box experiment. Consider a wall in which 
there are three holes $A$, $B$ and $C$. In front of the wall there is a particle 
source $Q$. Behind the wall there is a particle detector $D$. Both $Q$ and $D$ 
are equidistant from the three holes. Behind $C$ there is one other device; its 
purpose is to cause a phase shift by $\pi$. Particles emerging from the wall are 
thus preselected in a `state' $\ket{\Psi_1}$ proportional to 
$\ket{a}+\ket{b}+\ket{c}$, where $\ket{a}$, $\ket{b}$ and $\ket{c}$ represent the 
respective alternatives `particle goes through $A$,' `particle goes through $B$,' 
and `particle goes through $C$,' while detected particles are postselected in a 
`state' $\ket{\Psi_2}$ proportional to $\ket{a}+\ket{b}-\ket{c}$.

We will consider two possible intermediate measurements. First we place near 
$A$ a device $F_a$ that beeps whenever a particle passes through $A$. With 
the help of the ABL formula one finds that every particle of this particular
pre- and postselected ensemble $\cal E$ causes $F_a$ to beep with 
probability~1, as one may verify by calculating the probability with which a 
particle would be found passing through the union $B\cup C$ of $B$ and $C$:
$$
P_{ABL}(\subset B\cup C)\propto
\absosq{
\sandwich{\Psi_2}{{\bf P}_{\subset B\cup C}}{\Psi_1}
}=0,
$$
where ${\bf P}_{\subset B\cup C}=\ketbra bb+\ketbra cc$ projects on the 
subspace corresponding to the alternative `particle goes through $B\cup C$.' 
We obtain the same result by considering what would happen if $A$ were 
closed, or if all particles that make $F_a$ beep were removed from $\cal E$. 
The remaining particles are pre- and postselected in `states' proportional to 
$\ket{b}+\ket{c}$ and $\ket{b}-\ket{c}$, respectively, and these `states' are 
orthogonal. The result is an empty ensemble: if $A$ were closed, no particle 
would arrive at $D$. Does this warrant the conclusion that all particles 
belonging to $\cal E$ pass through $A$?

Let us instead place near $B$ a device $F_b$ that beeps whenever a particle 
passes through $B$. Considering the invariance of $\ket{\Psi_1}$ and 
$\ket{\Psi_2}$ under interchange of $\ket{a}$ and $\ket{b}$, one is not surprised 
to find that the ensemble $\cal E$ would be empty if the particles causing 
$F_b$ to beep were removed. Hence if the conclusion that all particles 
belonging to $\cal E$ pass through $A$ is warranted, so is the conclusion that 
the same particles also pass through $B$. If these `conclusions' were 
legitimate, they would make nonsense of the very concept of localization. 
Therefore we are forced to conclude instead that an ABL probability equal to 1 
does {\it not} warrant the existence of a corresponding element of reality (in the 
straightforward, ontological sense). Taken in this sense, (ER2) is false.

It pays to investigate further. We are in fact dealing with four different 
experimental arrangements: (i)~there is no beeper, (ii)~$F_a$ is the only 
beeper in place, (iii)~$F_b$ is the only beeper in place, (iv)~both $F_a$ and 
$F_b$ are in place. The first arrangement permits no legitimate inference 
concerning the hole taken by a particle. Assuming that $F_a$ is 100\% 
efficient, the second arrangement guarantees that one of two inferences is 
warranted: `the particle goes through $A$' (in case $F_a$ beeps) or `the 
particle goes through $B\cup C$' (in case $F_a$ fails to beep). Assuming that 
$F_b$ is equally efficient, the third arrangement likewise guarantees that one 
of two inferences is warranted: `the particle goes through $B$' or `the particle 
goes through $A\cup C$.' The fourth arrangement, finally, guarantees that one 
of three inferences is warranted: `the particle goes through $A$' (in case $F_a$ 
beeps), `the particle goes through $B$' (in case $F_b$ beeps), and `the particle 
goes through $C$' (in case neither $F_a$ nor $F_b$ beeps). The following 
counterfactuals are therefore true: (i)~If $F_a$ were in place, either it would 
beep and the particle would go through $A$, or it would fail to beep and the 
particle would go through $B\cup C$. (ii)~If $F_b$ were in place, either it would 
beep and the particle would go through $B$, or it would fail to beep and the 
particle would go through $A\cup C$. (iii)~If both $F_a$ and $F_b$ were in 
place, one of the following three conjunctions would be true: $F_a$ beeps and 
the particle goes through $A$; $F_b$ beeps and the particle goes through $B$; 
neither beeper beeps and the particle goes through $C$.

If we confine the discussion to particles that are emitted by $Q$ and detected 
by $D$, then the following counterfactuals are true: If $F_a$ but not $F_b$ 
were present, the alternatives represented by $\ket{b}$ and $\ket{c}$ would 
interfere with each other but not with the alternative represented by $\ket{a}$; 
as a consequence, they would interfere destructively; therefore $F_a$ would 
beep and the particle would go through $A$. By the same token, if $F_b$ were 
the only beeper present, the alternatives represented by $\ket{a}$ and $\ket{c}$ 
would interfere destructively, $F_b$ would beep, and the particle would go 
through $B$. Finally, if both beepers were present, no interference would take 
place; each particle would go through a particular hole, but not all particles 
would go through the same hole. To my mind, these counterfactuals are 
unobjectionable. One does not have to delve into the general philosophy of 
counterfactuals to see that they are true. I concur with Vaidman (1999) that 
quantum counterfactuals are unambiguous. Quantum counterfactuals are 
statements about possible worlds in which the outcomes of all measurements 
but one are the same as in the actual world. The remaining measurement is 
performed in a number of possible worlds (the number depends on the range of 
possible values) but not in the actual world.

The three-box (or three-hole) experiment demonstrates that position 
probabilities cannot be assigned independently of experimental arrangements. 
More specifically, they cannot be assigned without specifying a set of 
experimentally distinguishable alternatives. A position probability of~1 depends 
not only on the way the particle is `prepared' and `retropared' but also on the 
set $L$ of alternative locations that can be experimentally distinguished. If 
$L=\{A,B\cup C\}$, the particle is certain to be found in (or going through) $A$, 
but the inference of an element of reality `the particle went through $A$' is 
warranted only if the members of $L$ are actually distinguished (that is, only if 
the corresponding experiment is actually performed).

It follows that the position of a particle is an extrinsic property. By an {\it 
extrinsic} property $p$ of $S$ I mean a property of $S$ that is undefined unless 
either the truth or the falsity of the proposition ${\bf p}=$~`$S$~is~$p$' can be 
inferred from what happens or is the case in the `rest of the world' ${\cal W}-S$. 
The position of a particle is undefined unless there is a specific set $\{R_i\}$ of 
alternative locations, and unless there is a matter of fact about the particular 
location $R_j$ at which the particle is, or has been, present. (By `a matter of 
fact about the particular location $R_j$' I mean an actual event or an actual 
state of affairs from which that location can be inferred.\fnote{%
Examples of actual events are the click of a Geiger counter or the deflection of 
a pointer needle. An actual state of affairs is expressed, for instance, by the 
statement `The needle points to the left.' Can such events and states of affairs 
be defined in quantum-mechanical terms? See below.}%
) Positions are {\it defined} in terms of position-indicating matters of fact. They 
`dangle' from actual events or actual states of affairs. And if it is true that 
`[t]here is nothing in quantum theory making it applicable to three atoms and 
inapplicable to $10^{23}$' (Peres and Zurek, 1982), this must be as true of 
footballs and cats as it is of particles and atoms. The positions of things {\it are} 
what matters of fact imply concerning the positions of things.

If this is correct then (ER1) is as false as (ER2). In particular, the `sufficiency 
condition' (ER1) is {\it not} sufficient for the presence of a material object $O$ 
in a region of space $R$. The condition that is both necessary and sufficient for 
the presence of $O$ in $R$, is the existence of a matter of fact that indicates 
$O$'s presence in $R$. If there isn't any such matter of fact (now or anytime 
past or future), and if there also isn't any matter of fact that indicates $O$'s 
absence from $R$, then the sentence `$O$ is in $R$' is neither true nor false 
but meaningless, and $O$'s position with respect to $R$ (inside or outside) is 
undefined.

\section{\pcer}Probabilities, Conditonals, Elements of Reality, and the 
Measurement Problem

In the previous section I made use of the ABL probabilities associated with 
different versions of Vaidman's three-box experiment to show that positions are 
extrinsic, and that, consequently, both (ER1) and (ER2) are false. The validity 
of arguments based on time-symmetric quantum counterfactuals might be 
challenged. In the present section I therefore show without recourse to ABL 
probabilities that positions are extrinsic and that (ER1) is false. In the following 
section I shall establish the cogency of the argument of the previous section by 
showing that the proper condition for the truth of a quantum counterfactual is 
$P_{ABL}=1$. (It is readily verified that $P_B=1$ is sufficient but not necessary 
for this condition to be met.)

Consider two perfect detectors $D_1$ and $D_2$ whose respective (disjoint) 
sensitive regions are $R_1$ and $R_2$. If the support of the (normalized) wave 
function associated with the (center-of-mass) position of an object $O$ is 
neither wholly inside $R_1$ nor wholly inside $R_2$, nothing necessitates the 
detection of $O$ by $D_1$, and nothing necessitates the detection of $O$ by 
$D_2$. But if the wave function vanishes outside $R_1\cup R_2$, the 
probabilities for either of the detectors to click add up to~1, so either of the 
detectors is certain to click. Two perfect detectors with sensitive regions $R_1$ 
and $R_2$ constitute one perfect detector $D$ with sensitive region $R_1\cup 
R_2$. But how can it be certain that one detector will click when individually 
neither detector is certain to click? What could cause $D$ to click while causing 
neither $D_1$ nor $D_2$ to click?

That two perfect detectors with disjoint sensitive regions constitute one perfect 
detector for the union of the two regions forms part of the {\it definition} of what 
we mean by a perfect detector. By definition, a perfect detector clicks when the 
quantum-mechanical probability for it to click is~1. $D$ is certain to click 
because the probabilities for either of the two detectors to click add up to~1. 
Hence the question of what {\it causes} $D$ to click does not arise. Perfect 
detectors are theoretical constructs that by definition behave in a certain way. If 
real detectors would behave in the same way, it would be proper to inquire why 
they behave in this way. But real detectors are not perfect and do not behave in 
this way. A real detector is not certain to click when the corresponding 
quantum-mechanical probability is~1. Hence the question of what causes a real 
detector to click does not arise. Nothing causes a real detector to click.

What I aim at in this paper is an interpretation of quantum mechanics that takes 
standard quantum mechanics to be fundamental and complete. My claim that 
nothing causes a real detector to click is based (i)~on this assumption and 
(ii)~on the observation that the efficiency of a real detector cannot be 
accounted for in quantum-mechanical terms. All quantum-mechanical 
probability assignments are relative to {\it perfect} detectors. If quantum 
mechanics predicts that $D_1$ will click with a probability of $1/2$, it does {\it 
not} predict that a {\it real} detector will click in 50\% of all runs of the actual 
experiment. What it predicts is that $D_1$ will click in 50\% of {\it those} runs of 
the experiment in which either $D_1$ or $D_2$ clicks. Quantum mechanics 
has nothing to say about the percentage of runs in which no counter clicks (that 
is, is tells us nothing about the efficiency of $D$, or of any other real detector 
for that matter). If quantum mechanics predicts that $D_1$ will click with 
probability~1, it accounts for the fact that whenever either $D_1$ or $D_2$ 
clicks, it is $D_1$ that clicks. It does {\it not} account for the clicking of either 
$D_1$ or $D_2$. Where quantum mechanics is concerned, nothing causes the 
clicking. And if quantum mechanics is as fundamental and complete as is here 
assumed, then this is true without qualification: nothing causes the 
clicking.\fnote{%
It is well known that all actually existing detectors are less than perfect. On the 
other hand, there is no (obvious) theoretical limit to the efficiency of a real 
detector. It might one day be possible to build a detector with an efficiency 
arbitrarily close to 100\%. However, unless the efficiency of detectors is exactly 
100\%, it remains impossible to interpret a `preparation' that warrants assigning 
probability~1 as {\it causing} a detector to click. If the preparation is to be a 
sufficient reason for the click, the detector must {\it always} click (that is, it must 
be perfect). What if it were possible to build perfect detectors? We could then 
speak of the preparation as the cause of the click, but if quantum mechanics is 
fundamental and complete, it would still be impossible to explain how the 
preparation causes the click: {\it ex hypothesi}, no underlying mechanism 
exists. The perfect correlation between preparation and click would have to be 
accepted as a brute fact. So would the fact that either $D_1$ or $D_2$ clicks 
when neither of them is certain to click. Causality would be just another name 
for such correlations, not an explanation.}

Quantum mechanics assigns probabilities (whether Born or ABL) to alternative 
events (e.g., deflection of the pointer needle to the left or to the right) or to 
alternative states of affairs (e.g., the needle's pointing left or right). Implicit in 
every normalized distribution of probabilities over a specified set of alternative 
events or states of affairs is the assumption that exactly one of the specified 
alternatives happens or obtains. If we assign normalized probabilities to a set of 
counterfactuals, we still assume (counterfactually) that exactly one of the 
counterfactuals is true. In other words, if we assign probabilities to the possible 
results of an unperformed measurement, we still assume that the 
measurement, if it had been performed, would have yielded a definite result. 
Like all (normalized) probabilities, the probabilities assigned by quantum 
mechanics are assigned to mutually exclusive and jointly exhaustive 
possibilities, and they are assigned {\it on the supposition} that exactly one 
possibility is, or would have been, a fact. Even the predictions of the standard 
version of standard quantum mechanics therefore are {\it conditionals}. 
Everything this version tells us conforms to the following pattern: {\it If} there is 
going to be a matter of fact about the alternative taken (from a specific range of 
alternatives), {\it then} such and such are the Born probabilities with which that 
matter of fact will indicate this or that alternative.

It is important to understand that quantum mechanics never allows us to predict 
{\it that} there will be such a matter of fact, unconditionally. If the Born 
probability of a particular event $F$ is~1, we are not entitled to predict that $F$ 
will happen. What we are entitled to infer is only this: Given that one of a 
specified set of events will happen, and given that $F$ is an element of this set, 
the event that will happen is $F$. In order to get from a true conditional to an 
element of reality, a condition has to be met: a measurement must be 
successfully performed, there must be a matter of fact about the value of an 
observable, one of a specific set of alternative property-indicating events or 
states of affairs must happen or obtain. Quantum mechanics does not predict 
that a measurement will take place, nor the time at which one will take place, 
nor does it specify the conditions in which one will take place. It requires us to 
{\it assume} that one will take place, for it is on this assumption that its 
probability assignments are founded.

It follows that (ER1) is false. A Born probability equal to~1 is equivalent to a 
conditional $c$. The inference of a corresponding element of reality is 
warranted only if the condition laid down by $c$ is actually met.

It also follows that positions are extrinsic. The condition laid down by $c$ is the 
existence of a matter of fact about the value taken by some observable. If this 
observable has for its spectrum a set $\{R_i\}$ of mutually disjoint regions of 
space, if $R$ is an element of $\{R_i\}$, and if the Born probability associated 
with $R$ is~1, then $O$ is inside $R$ just in case there is a matter of fact 
about the particular element of $\{R_i\}$ that contains $O$.

I conclude this section with a few remarks concerning the so-called 
measurement problem. First some basic facts. Quantum mechanics represents 
the possible values $q^k_i$ of all observables $Q^k$ as projection operators 
${\bf P}_{Q^k=q^k_i}$ on some Hilbert space $\cal H$. The projection 
operators that jointly represent the range of possible values of a given 
observable are mutually orthogonal. If one defines the `state' of a system as a 
{\it probability measure} on the projection operators on $\cal H$ (Cassinello and
S\'anchez-G\'omez, 1996; Jauch, 1968, p. 94) resulting from a {\it preparation} 
of the system (Jauch, 1968, p. 92), one finds (Cassinello and
S\'anchez-G\'omez, 1996; Jauch, 1968, p. 132) that every such probability 
measure has the form $P({\bf P})=\hbox{Tr}({\bf WP})$, where $\bf W$ is a 
unique density operator (that is, a unique self-adjoint, positive operator 
satisfying $\hbox{Tr}({\bf W})=1$ and ${\bf W}^2\leq{\bf W}$). [The trace 
$\hbox{Tr}({\bf X})$ is the sum $\sum_i\sandwich{i}{{\bf X}}{i}$, where 
$\{\ket{i}\}$ is any orthonormal basis in $\cal H$.] If ${\bf W}^2(t)={\bf W}(t)$, 
${\bf W}(t)$ projects on a one-dimensional subspace of $\cal H$ and thus is 
equivalent -- apart from an irrelevant phase factor -- to a `state' vector 
$\ket{\Psi(t)}$ or a wave function $\Psi(x,t)$, $x$ being any point in the 
system's configuration space. In this case one retrieves the Born formula~(1).

The quantum-mechanical `state' vector (or the wave function, or the density 
operator) thus is essentially a probability measure on the projection operators 
on $\cal H$, specifying probability distributions over all sets of mutually 
orthogonal subspaces of $\cal H$. Hence the $t$ that appears in the `states' 
${\bf W}(t)$, $\ket{\Psi(t)}$, and $\Psi(x,t)$ has the same significance as the $t$ 
that appears in the time-dependent probabilities $P_B(q^k_i,t)$. Now recall that 
quantum mechanics predicts neither that a measurement will take place nor the 
time at which one will take place. It requires us to {\it assume} that a 
measurement will take place {\it at a specified time}. The time-dependence of 
the `state' vector therefore is a dependence on the {\it specified} time at which 
a {\it specified} observable (with a {\it specified} range of values) is measured 
either in the actual world or in a set of possible worlds. It is {\it not} the
time-dependence of a state of affairs that evolves in time.

On the supposition that $\ket{\Psi(t)}$ represents a state of affairs that evolves 
in time (so that at every time $t$ a state of affairs $\ket{\Psi(t)}$ obtains), one 
needs to explain what brings about the real or apparent discontinuous transition 
from a state of affairs represented by a ket of the form $\ket{\Psi(t)}=\sum_i 
a_i(t)\ket{a_i}$ to the state of affairs represented by one of the kets $\ket{a_i}$. 
This is the measurement problem. It is a pseudoproblem because a collection 
of time-dependent probabilities is not a state of affairs that evolves in time. The 
probability for something to happen at the time $t$ does not exist at $t$, any 
more than the probability for something to be located in $R$ exists in $R$.

The probabilities $P_B(q^k_i,t)$ are determined by the relevant matters of fact 
about the properties possessed by a physical system at or before a certain time 
$t_0$. In the special case of a complete measurement performed at $t_0$, 
they are given by the Born formula
$$
P_B(q^k_i,t)=|\sandwich{\Psi(t)}{{\bf P}_{Q^k=q^k_i}}{\Psi(t)}|\qquad
\hbox{for $t\geq t_0$},\eqno{(3)}
$$
where $\ket{\Psi(t)}=U(t-t_0)\ket{\Psi(t_0)}$. $\ket{\Psi(t_0)}$ is the `state' 
`prepared' by the measurement at $t_0$ (that is, it represents the properties 
possessed by the system at $t_0$). $U(t-t_0)$ is the unitary operator that 
governs the time-dependence of quantum-mechanical probabilities (often 
misleadingly referred to as the `time evolution operator'). And $t$ is the {\it 
stipulated} time at which the next measurement is performed, either actually or 
counterfactually.

Thus all that a superposition of the form $\ket{\Psi(t)}=\sum_i a_i(t)\ket{a_i}$ 
tells us, is this: {\it If} there is a matter of fact from which one can infer the 
particular property (from the set of properties represented by the kets 
$\ket{a_i}$) that is actually possessed by the system at the stipulated time $t$, 
then the prior probability that this matter of fact indicates the property 
represented by $\ket{a_i}$ is $\absosq{a_i}$. It is self-evident that if there {\it is} 
such a matter of fact, and if this matter of fact is taken into account, the correct 
basis for further conditional inferences is not $\ket{\Psi(t)}$ but one of the kets 
$\ket{a_i}$. This obvious truism is the entire content of the so-called projection 
postulate (L\"uders, 1951; von Neumann, 1955). By the same token, all that an 
entangled `state' of the form $\sum_i a_i(t)\ket{b_i}\otimes\ket{a_i}$ tells us, is 
this: {\it If} there are two matters of fact, one indicating which of the properties 
represented by the kets $\ket{a_i}$ is possessed by the first system, and 
another indicating which of the properties represented by the kets $\ket{b_i}$ is 
possessed by the second system, then the two matters of fact together indicate 
$\ket{a_i}$ and $\ket{b_i}$ with probability $\absosq{a_i}$, and they indicate 
$\ket{a_i}$ and $\ket{b_j}$ ($j\neq i$) with probability~0. But if there {\it is} any 
such matter of fact, these entangled probabilities are based on an incomplete 
set of facts and are therefore subjective (that is, they reflect our ignorance of 
some relevant fact). All that ever gets {\it objectively} entangled is {\it 
counterfactuals}.

\section{\opretro}Objective Probabilities, Retrocausation, and the Arrow of Time

What most strikingly distinguishes quantum physics from classical physics is 
the existence of objective probabilities. In a classical world there are no
(nontrivial) objective probabilities: the probability of dealing an ace is not 
$1/13$ but either $1$ or $0$, depending on whether or not an ace is top card. 
Objective probabilities have nothing to do with ignorance; there is nothing (that 
is, no actual state of affairs, no actually possessed property, no actually 
obtained measurement result) for us to be ignorant of. Then what is it that {\it 
has} an objective probability? What are objective probabilities distributed over? 
The obvious answer is: counterfactuals. Only a contrary-to-fact conditional can 
be assigned an objective probability. Objective probabilities are distributed over 
the possible results of {\it unperformed} measurements. Objective probabilities 
are objective in the sense that they are not subjective, and they are not 
subjective because they would be so only if the corresponding measurements 
were performed. In short, objective probabilities are probabilities that are {\it 
counterfactually subjective}.

Probabilities can be objective only if they are based on a complete set of facts. 
Otherwise they are subjective: they reflect our ignorance of some of the 
relevant facts. Born probabilities are in general calculated on the basis of an 
incomplete set of facts; they take into account the relevant past matters of fact 
but ignore the relevant future matters of fact. Born probabilities are objective 
only if there are no relevant future matters of fact. Thus they cannot be 
objective if any one of the measurements to the possible results of which they 
are assigned, is actually performed. This is equally true of ABL probabilities: if 
one of the measurements to the possible results of which they are assigned, is 
actually performed, they too are calculated on the basis of an incomplete set of 
facts. They take into account all revelant matters of fact except the result of the 
actually performed measurement. On the other hand, if none of these 
measurements is actually performed, ABL probabilities take into account all 
relevant matters of fact and are therefore objective.

Thus probabilities are objective only if they are distributed over alternative 
properties or values none of which are actually possessed, and only if they are 
based on all relevant events or states of affairs, including those that are yet to 
occur or obtain. In general the objective probabilities associated with
contrary-to-fact conditionals depend also on events that haven't yet happened 
or states of affairs that are yet to obtain. Hence some kind of retroactive 
causality appears to be at work. This necessitates a few remarks concerning 
causality and the apparent `flow' of time.

But first let us note that nothing entails the existence of time-reversed causal 
connections between {\it actual} events and/or states of affairs. To take a 
concrete example, suppose that at $t_1$ the $x$ component $\sigma_x$ of the 
spin of an electron is measured, that at $t_2>t_1$ $\sigma_y$ is measured, 
and that the respective results are $\uparrow_x$ and $\uparrow_y$. Then a 
measurement of $\sigma_x$ would have yielded $\uparrow_x$ if it had been 
performed at an intermediate time $t_m$, and a measurement of $\sigma_y$ 
would have yielded $\uparrow_y$ if it had been performed instead. What would 
have happened at $t_m$ depends not only on what happens at $t_1$ but also 
on what happens at $t_2$. But if either $\sigma_x$ or $\sigma_y$ is actually 
measured at $t_m$ (other things being equal), nothing compels us to take the 
view that $\uparrow_y$ was obtained at $t_m$ {\it because} the same result 
was obtained at $t_2$. We can stick to the idea that causes precede their 
effects, according to which $\uparrow_y$ was obtained at $t_2$ {\it because} 
the same result was obtained at $t_m$. The point, however, is that nothing in 
the physics {\it prevents} us from taking the opposite view. The distinction we 
make between a cause and its effect is based on the apparent `motion' of our 
location in time -- the present moment -- toward the future. This special location 
and its apparent `motion' are as extraneous to physics as are our location and 
motion in space (Price, 1996). Equally extraneous, therefore, is the distinction 
between causes and effects.

Physics deals with correlations between actual events or states of affairs, 
classical physics with deterministic correlations, quantum physics with 
statistical ones. Classical physics allows us to explain the deterministic 
correlations (abstracted from what appear to be universal regularities) in terms 
of causal links between individual events. And for some reason to be explained 
presently, we identify the earlier of two diachronically correlated events as the 
cause and the later as the effect. The time symmetry of the classical laws of 
motion, however, makes it equally possible to take the opposite view, 
according to which the later event is the cause and the earlier event the effect. 
In a deterministic world, the state of affairs at any time $t$ determines the state 
of affairs at any other time $t'$, irrespective of the temporal order of $t$ and 
$t'$. The belief in a time-asymmetric {\it physical} causality is nothing but an 
animistic projection of the perspective of a conscious agent into the inanimate 
world, as I proceed to show.

I conceive of myself as a causal agent with a certain freedom of choice. But I 
cannot conceive of my choice as exerting a causal influence on anything that I 
knew, or could have known, at the time $t_c$ of my choice. I can conceive of 
my choice as causally determining only such events or states of affairs as are 
unknowable to me at $t_c$. On the simplest account, what I knew or could 
have known at $t_c$ is everything that happened before $t_c$. And what is 
unknowable to me at $t_c$ is everything that will happen thereafter. This is the 
reason why we tend to believe that we can causally influence the future but not 
the past. And this constraint on {\it our} (real or imagined) causal efficacy is 
what we impose, without justification, both on the deterministic world of 
classical physics and on the indeterministic world of quantum physics.

In my goal-directed activities I exploit the time-symmetric laws of physics. 
When I kick a football with the intention of scoring a goal, I make (implicit or 
explicit) use of my knowledge of the time-symmetric law that governs the ball's 
trajectory. But my thinking of the kick as the cause and of the scored goal as its 
effect has nothing to do with the underlying physics. It has everything to do with 
my self-perception as an agent and my successive experience of the world. 
The time-asymmetric causality of a conscious agent in a successively 
experienced world rides piggyback on the symmetric determinisms of the 
physical world, and in general it rides into the future because in general the 
future is what is unknowable to us. But it may also ride into the past. Three 
factors account for this possibility.

First, as I said, the underlying physics is time-symmetric. If we ignore the 
strange case of the neutral kaon (which doesn't appear to be relevant to the 
interpretation of quantum mechanics), this is as true of quantum physics as it is 
of classical physics. If the standard formulation of quantum physics is 
asymmetric with respect to time, it is because we think (again without 
justification) that a measurement does more than yield a particular result. We 
tend to think that it also {\it prepares} a state of affairs which evolves toward the 
future. But if this is a consistent way of thinking -- it is {\it not} (Mohrhoff, 1999) 
-- then it is equally consistent to think that a measurement `retropares' a state 
of affairs that evolves toward the past, as Aharonov, Bergmann, and Lebowitz 
(1964) have shown.

Second, what matters is what can be known. If I could know the future, I could 
not conceive of it as causally dependent on my present choice. In fact, if I 
could (in principle) know both the past and the future, I could not conceive of 
myself as an agent. I can conceive of my choice as causally determining the 
future precisely because I cannot know the future. This has nothing to do with 
the truism that the future does not (yet) exist. Even if the future in some way 
`already' exists, it can in part be determined by my present choice, provided I 
cannot know it at the time of my choice. By the same token, a past state of 
affairs can be determined by my present choice, provided I cannot know that 
state of affairs before the choice is made.

There are two possible reasons why a state of affairs $F$ cannot be known to 
me at a given time $t$: (i)~$F$ may obtain only after $t$; (ii)~at $t$ there may 
as yet exist no matter of fact from which $F$ can be inferred. This takes us to 
the last of the three factors which account for the possibility of retrocausation: 
the contingent properties of physical systems are extrinsic. By a {\it contingent} 
property I mean a property that may or may not be possessed by a given 
system at a given time. For example, being inside a given region of space and 
having a spin component of $+\hbar/2$ along a given axis are contingent 
properties of electrons.

Properties that can be retrocausally determined by the choice of an 
experimenter, cannot be intrinsic. [A property $p$ of a physical system $S$ is 
{\it intrinsic} iff the proposition ${\bf p}=$~`$S$~is~$p$' is `of itself' (that is, 
unconditionally) either true or false at any time.] If $p$ is an extrinsic property 
of $S$, the respective criteria for the truth and the falsity of the proposition 
${\bf p}=$~`$S$~is~$p$' are to be sought in the `rest of the world' ${\cal W}-S$, 
and it is possible that neither criterion is satisfied, in which case $\bf p$ is 
neither true nor false but meaningless. It is also possible that each criterion 
consists in an event that may occur only after the time to which $\bf p$ refers. 
If this event is to some extent determined by an experimenter's choice, 
retrocausation is at work. On the other hand, if $p$ is an intrinsic property of 
$S$, $\bf p$ has a truth value (`true' or `false') independently of what happens 
in ${\cal W}-S$, so {\it a fortiori} it has a truth value independently of what 
happens there after the time $t$ to which $\bf p$ refers. There is then no 
reason why the truth value of $\bf p$ should be unknowable until some time 
$t'>t$. In principle it is knowable at $t$, and therefore we cannot (or at any rate, 
need not) conceive of it as being to some extent determined by the 
experimenter's choice at $t'$.

A paradigm case of retrocausation at work (Mohrhoff, 1999) is the experiment 
of Englert, Scully, and Walther (1994; Scully, Englert, and Walther, 1991). This 
experiment permits the experimenters to choose between (i)~measuring the 
phase relation with which a given atom emerges coherently from (the union of) 
two slits and (ii)~determining the particular slit from which the atom emerges. 
The experimenters can exert this choice after the atom has emerged from the 
slit plate and even after it has hit the screen. By choosing to create a matter of 
fact about the slit taken by the atom, they retroactively cause the atom to have 
passed through a particular slit. By choosing instead to create a matter of fact 
about the atom's phase relation, they retroactively cause the atom to have 
emerged with a definite phase relation. The retrocausal efficacy of their choice 
rests on the three factors listed above (in different order): (i)~The four 
propositions ${\bf a}_1=$~``the atom went through the first slit,'' ${\bf 
a}_2=$~``the atom went through the second slit,'' ${\bf a}_+=$~``the atom 
emerged from the slits in phase,'' and ${\bf a}_-=$~``the atom emerged from 
the slits out of phase'' affirm {\it extrinsic} properties. (ii)~There exist
{\it time-symmetric} correlations between the atom's possible properties at the 
time of its passing the slit plate and the possible results of two mutually 
exclusive experiments that can be performed at a later time. (iii)~The result of 
the actually performed experiment is the first (earliest) matter of fact about 
either the particular slit taken by the atom or the phase relation with which the 
atom emerged from the slits. Before they made their choice, the experimenters 
could not possibly have {\it known} the slit from which, or the phase relation 
with which, the atom emerged.

Probabilities, I said, can be objective only if they are based on all relevant 
matters of fact, including those still in the future. We are now in a position to 
see clearly why it should be so. Our distinction between the past, the present, 
and the future has nothing to do with physics. Physics knows nothing of the 
experiential {\it now} (the special moment at which the world has the technicolor 
reality it has in consciousness), nor does it know anything of the difference 
between what happened before {\it now} (the past) and what will happen after 
{\it now} (the future).

\longsection{\qmcc}The World According to Quantum Mechanics: 
Fundamentally Inexplicable

Correlations Between Fundamentally Inexplicable Events

It is commonly believed that it is the business of quantum mechanics to 
account for the occurrence/existence of actual events or states of affairs.
Environment-induced superselection (Joos and Zeh, 1985; Zurek, 1981, 1982), 
decoherent histories (Gell-Mann and Hartle, 1990; Griffiths, 1984; Omn\`es, 
1992), quantum state diffusion (Gisin and Percival, 1992; Percival 1994), and 
spontaneous collapse (Ghirardi, Rimini and Weber, 1986; Pearle, 1989) are 
just some of the strategies that have been adopted with a view to explaining 
the emergence of `classicality.' Whatever is achieved by these interesting 
endeavors, they miss this crucial point: quantum mechanics only takes us from 
facts to probabilities of possible facts. The question of how it is that exactly one 
possibility is realized must not be asked of a formalism that serves to assign 
probabilities on the implicit {\it assumption} that exactly one of a specified set 
of possibilities {\it is} realized. Even the step from probability~1 to factuality 
crosses a gulf that quantum mechanics cannot bridge. Quantum mechanics 
can tell us that $O$ is certain to be found in $R$ {\it given that} there is a 
matter of fact about its presence or otherwise in $R$, but only the actual matter 
of fact warrants the inference that $O$ is in $R$.

Quantum mechanics does not predict that a measurement will take place, nor 
the time at which one will take place, nor does it specify the conditions in which 
one will take place. And if quantum mechanics is as fundamental as I presume 
it is, {\it nothing} allows us to predict that or when a measurement takes place, 
or to specify conditions in which one is certain to take place, for there is nothing 
that {\it causes} a measurement to take place. In other words, a matter of fact 
about the value of an observable is a causal primary. A {\it causal primary} is 
an event or state of affairs the occurrence or existence of which is not 
necessitated by any cause, antecedent or otherwise.

I do not mean to say that in general nothing causes a measurement to yield this 
rather than that particular value. Unless one postulates hidden variables, this is 
a triviality. What I mean to say is that nothing ever causes a measurement to 
take place. Measurements (and in clear this means detection events) are 
causal primaries. No detector is 100\% efficient. Using similar detectors in 
series, it is easy enough to experimentally establish a detector's (approximate) 
likelihood to click when the corresponding Born probability is~1, but of this 
likelihood no theoretical account is possible.\fnote{%
There are two kinds of probability, the probability {\it that} a detector will 
respond (rather than not respond), and the probability that this (rather than any 
other) detector will respond {\it given} that exactly one detector will respond. 
The former probability cannot be calculated using the quantum formalism (nor, 
if quantum mechanics is fundamental and complete, any other formalism). One 
can of course analyze the efficiency of, say, a Geiger counter into the 
efficiencies of its `component detectors' (the ionization cross sections of the 
ionizable targets it contains), but the efficiencies of the `elementary detectors' 
cannot be analyzed any further. This entails that a fundamental coupling 
constant such as the fine structure constant cannot be calculated from `first 
principles;' it can only be gleaned from the experimental data.}
{\it A fortiori}, no theoretical account is possible of why or when a detector is 
certain to click. It never is.

Quantum physics thus is concerned with correlations between events or states 
of affairs that are uncaused and therefore fundamentally inexplicable. As 
physicists we are not likely to take kindly to this conclusion, which may account 
for the blind spot behind which its inevitability has been hidden so long. But we 
certainly {\it are} at a loss when it comes to accounting for the world of definite 
occurrences. Recently Mermin (1998) advocated an interpretation of the 
formalism of standard quantum mechanics according to which ``[c]orrelations 
have physical reality; that which they correlate, does not.'' He does not claim 
that there are no correlata, only that they are not part of {\it physical} reality. 
The correlated events belong to a larger reality which includes consciousness 
and which lies outside the scope of physics. Thus Mermin agrees that, where 
physics is concerned, the correlata {\it are} fundamentally inexplicable.

The idea that the correlata are conscious perceptions (Lockwood, 1989; Page, 
1996), or beliefs (Albert, 1992), or knowings (Stapp, 1993) has a respectable 
pedigree (London and Bauer, 1939; von Neumann, 1955). If one thinks of the 
state vector as representing a state of affairs that evolves in time, one needs 
something that is `more actual' than the state vector -- something that bestows 
`a higher degree of actuality' than does the state vector -- to explain why every 
successful measurement has exactly one result, or why measurements are 
possible at all. This is the spurious measurement problem all over again. It is 
spurious because the state vector does {\it not} represent an evolving state of 
affairs. If we were to relinquish this unwarranted notion, we would not need two 
kinds of reality to make sense of quantum mechanics, such as a physical reality 
and a reality that includes consciousness (Mermin, 1998), or a potential reality 
and an actual reality (Heisenberg, 1958; Popper, 1982; Shimony, 1978, 1989), 
or a mind-constructed `empirical' reality and a mind-independent `veiled' reality 
(d'Espagnat, 1995), or an unrecorded `smoky dragon' reality and an irreversibly 
recorded reality (Wheeler, 1983). We could confine ourselves to talking about 
events that are causal primaries, the inferences that are warranted by such 
events, the correlations between such events or such inferences, and the 
further inferences that are warranted by these correlations.

I do not deny that there is a larger reality that includes consciousness and that 
lies outside the scope of physics. What I maintain is that the interpretative 
problems concerning quantum mechanics can be solved without appealing to 
any larger reality, and that such an appeal does not help solving those 
problems because it is neither necessary nor possible to account for the 
occurrence of a causal primary. Theoretical physics is partly mathematics and 
partly semantics. The semantic task is to name the fundamental 
epistemological and/or ontological entities and/or relations represented by the 
symbols of the formalism. I cannot think of a more satisfactory choice of a 
basic (and therefore not further explicable) ontological entity for a physical 
theory than a causal primary -- something that is inexplicable {\it by definition}. 
Ever since the seminal paper by Einstein, Podolsky, and Rosen (1935), it has 
been argued that quantum mechanics is incomplete (Bell, 1966; Ford and 
Mantica, 1992; Lockwood, 1989; Primas, 1990). In point of fact, no theory can 
be more complete (with regard to its subject matter) than one that accounts for 
everything (within its subject matter) but what is inexplicable by definition. If 
there is anything that is incomplete, it is reality itself (that is, reality is 
incomplete relative to our {\it description} of it, which is `overcomplete') -- but 
I'm getting ahead of myself.

Because the occurrence/existence of actual events or states of affairs is 
presupposed by the formalism, locutions such as `actual event,' `actual state of 
affairs,' `matter of fact' cannot even be {\it defined} within the formalism. This 
conclusion too is unlikely to be popular with physicists, who naturally prefer to 
define their concepts in terms of the mathematical formalism they use. Einstein 
spent the last thirty years of his life trying (in vain) to get rid of field sources -- 
those entities that have the insolence to be real by themselves rather than by 
courtesy of some equation (Pais, 1982). Small wonder if he resisted Bohr's 
insight that not even the {\it properties} of things can be defined in purely 
mathematical terms. But Bohr was right. If Bohr (1934, 1963) insisted on the 
necessity of describing quantum phenomena in terms of experimental 
arrangements, it was because he held that the properties of quantum systems 
are {\it defined} by the experimental arrangements in which they are displayed 
(d'Espagnat, 1976).

For `experimental arrangement' read: what matters of fact permit us to infer 
concerning the properties of a given system at a given time. The contingent 
properties of physical systems are defined in terms of the actual events or 
states of affairs from which they can be inferred. They `dangle' from what 
happens or is the case in the rest of the world. They cannot be defined in purely 
mathematical terms, for only intrinsic properties can be so defined. The scope 
of physics is not restricted to laboratory experiments. {\it Any} matter of fact 
that has a bearing on the properties of a physical system qualifies as a 
`measurement result.' What is relevant is the occurrence or existence of an 
event or state of affairs warranting the assertability of a statement of the form 
`$S$ is $p$ (at the time $t$),' irrespective of whether anyone is around to 
assert, or take cognizance, of that event or state of affairs, and irrespective of 
whether it has been anyone's intention to learn something about $S$.

The following picture emerges. The world is a mass of events that are causal 
primaries. Without any correlations between these events, it would be a total 
chaos. As it turns out, the uncaused events are strongly correlated. If we don't 
look too closely, they fall into neat patterns that admit of being thought of as 
persistent objects with definite and continuously evolving positions. Projecting 
our time-asymmetric agent-causality into the time-symmetric world of physics, 
we think of the positions possessed at later times as causally determined by the 
positions possessed at earlier times. If we look more closely, we find that 
positions aren't always attributable, and that those that are attributable aren't 
always predictable on the basis of past events. We discover that positions do 
not `dangle' from earlier positions by causal strings but instead `dangle' from 
position-defining events that are statistically correlated but (being causal 
primaries) are not causally connected. Quantum mechanics describes the 
correlations but does nothing to explain them. Not only the correlata but also 
the correlations are incapable of (causal) explanation. Causal explanations are 
confined to the familiar macroworld of deterministic processes and things that 
evolve in time. This macroworld with its causal links is something we project 
onto the correlations and their uncaused correlata, but the projection works only 
to the extent that the correlations are not manifestly probabilistic.\fnote{%
This is discussed in the last two sections.}
There are no causal processes more fundamental than the correlations and 
their correlata, processes that could in any manner account for the correlations 
or the correlata.

\section{\uinex}Spatial Nonseparability

The remaining interpretative task thus consists not in explaining the 
correlations but in understanding what they are trying to tell us about the world. 
Here I will confine myself to discussing some of the implications of the 
diachronic correlations (the correlations between results of measurements 
performed on the same system at different times).\fnote{%
The implications of the synchronic (EPR) correlations have been discussed 
elsewhere (Mohrhoff, submitted).}

Perhaps the first insight one gleans from the correlations is the existence of 
persisting {\it entities}. If the correlations did not permit us to speak of such 
entities, we could not think of the correlata as possessed properties, extrinsic or 
otherwise. Suppose that we perform a series of position measurements. And 
suppose that every position measurement yields exactly one result (that is, 
each time exactly one detector clicks). Then we are entitled to infer the 
existence of an entity $O$ which persists through time (if not for all time), to 
think of the clicks given off by the detectors as matters of fact about the 
successive positions of this entity, to think of the behavior of the detectors as 
position measurements, and to think of the detectors as detectors. (The lack of 
transtemporal identity among particles of the same type of course forbids us to 
extend to such particles the individuality of a fully `classical' entity.) The 
successive positions of $O$, however, are extrinsic: they are what can be 
inferred from the pattern of clicks. All that can be inferred concerning $O$'s 
positions at times at which no detector clicks, is counterfactual and 
probabilistic.\fnote{%
The detectors of the present scenario are assumed to be time-specific: a click 
not only indicates a position but also the time at which it is possessed.}
There is a persistent entity all right, but there is then no actually possessed 
position to go with it.

The next lesson to be learned from the correlations is that the positions of 
things are objectively indefinite or `fuzzy.' This does not mean that $O$ {\it has} 
as fuzzy position. It means that statements of the form `$O$ is in $R$ at $t$' 
are sometimes neither true nor false but meaningless. This possibility stands or 
falls with the extrinsic nature of positions and the existence of objective 
probabilities. Take the counterfactual `If there were a matter of fact about the 
slit taken by the atom, the atom would have taken the first slit.' We can assign 
to this counterfactual an {\it objective} probability iff the proposition `The atom 
went through the first slit' is neither true nor false but meaningless. The reason 
why this proposition {\it can} be meaningless is that positions are extrinsic. It {\it 
is} meaningless just in case there isn't any matter of fact about the slit taken by 
the atom.

If it is true that the atom went through the union of the slits (that is, if the atom 
was emitted on one side of the slit plate and detected on the other side), and if 
it is meaningless to say that the atom went through the first slit (in which case it 
is also meaningless to say that it went through the second slit), then the 
conceptual distinction we make between the two slits has no reality for the 
atom. If that distinction were real for the atom (that is, if the atom behaved as if 
the two slits were distinct), the atom could not behave as if it went -- as a 
whole, without being divided into distinct parts -- simultaneously through both 
slits. But (if quantum mechanics is fundamental and complete) this is what the 
atom does when interference fringes are observed.

Thus there are objects for which our conceptual distinction between mutually 
disjoint regions of space does not exist. It follows that the distinction between 
such regions cannot be real {\it per se} (that is, it cannot be an intrinsic property 
of the world). If it were real {\it per se}, the following would be the true: at any 
one time, for every finite region $R$, the world can be divided into things or 
parts that are situated inside $R$, and things or parts that are situated inside 
the complement $R'$ of $R$. The boundary of $R$ would demarcate an 
intrinsically distinct part of the world. But if this were the case, exactly one of 
the following three propositions would be true of every object $O$ at any given 
time: (i)~$O$ is situated wholly inside $R$; (ii)~$O$ is situated wholly inside 
$R'$; (iii)~$O$ has two parts, one situated wholly inside $R$ and one situated 
wholly inside $R'$. If there is anything that (standard) quantum mechanics is 
trying to tells us about the world, it is that for at least some objects all of these 
propositions are sometimes false.

It follows that the multiplicity and the distinctions inherent in our {\it 
mathematical concept} of space -- a transfinite set of triplets of real numbers -- 
are {\it not} intrinsic features of {\it physical} space. The notion that these 
features of our mathematical concept of space are intrinsic to physical space -- 
in other words, the notion that the world is {\it spatially separable} -- is a 
delusion. This notion is as inconsistent with quantum mechanics as the notion 
of absolute simultaneity is with special relativity. `Here' and `there' are not {\it 
per se} distinct. Reality is {\it fundamentally nonseparable}. Like the positions of 
things, spatial distinctions `dangle' from actual events or states of affairs. 
Reality is not built on a space that is differentiated the way our mathematical 
concept of space is differentiated. A description of the world that incorporates 
such a space -- and {\it a fortiori} every description that identifies `the points of 
space (or space-time)' as the carriers of physical properties -- is 
`overcomplete.' Reality is built on matters of fact, and the actually existing 
differences between `here' and `there' are the differences that can be inferred 
from matters of fact. In and of itself, physical space -- or the reality underlying it 
-- is undifferentiated, one.

\section{\mapicos}Macroscopic Objects

The extrinsic nature of positions appears to involve a twofold vicious regress. 
To adequately deal with it, I need to talk about macroscopic objects. A {\it 
macroscopic object} $M$ is an object that satisfies the following criterion: any 
factually warranted inference concerning the position of $M$ at any time $t$ is 
predictable (with certainty) on the basis of factually warranted inferences about 
the positions of $M$ at earlier times. (A factually warranted inference is an 
inference that is warranted by some matter of fact.) Thus, to the extent that 
they can be inferred from actual events, the successive positions of a 
macroscopic object evolve deterministically. This makes it possible to ignore 
the fact that the positions of macroscopic objects, like all actually possessed 
positions, depend for their existence on position-indicating events. We can treat 
the positions of macroscopic objects as intrinsic properties and assume that 
they follow definite and causally determined trajectories, without ever risking to 
be contradicted by an actual event.

I do not mean to say that the position of $M$ really {\it is} definite. Even the 
positions of macroscopic objects are fuzzy, albeit not manifestly so: the 
positional indefiniteness of $M$ does not evince itself through unpredictable
position-indicating events. Nor do I mean to say that the positions of 
macroscopic objects really {\it are} intrinsic. They too `dangle' from actual 
events. But they do so in a way that is predictable, that does not reveal any 
fuzziness. We may think of macroscopic objects as following definite 
trajectories, or we may think of them as following fuzzy trajectories. Since all 
matters of fact about their positions are predictable, it makes no difference: the 
fuzziness has no factual consequences. Classical behavior results when the 
factually warranted positions fuse into a not manifestly fuzzy trajectory. It has 
little to do with the `classical limit' in which the wave packet shrinks to a 
continuously moving point, for the wave packet (of whatever size) is a bundle 
of probabilities associated with time-dependent counterfactuals, not the actual 
trajectory of an object.\fnote{%
Good examples of how not to get from quantum to classical are the 
unsuccessful attempts to obtain the exponential decay law, which pertains to 
factually warranted inferences and is consistent with all experimental data, from 
the Schr\"odinger equation, which tells us how the probabilities associated with 
counterfactuals depend on time (Onley and Kumar, 1992; Singh and Whitaker, 
1982).}

By saying that matters of fact about the positions of macroscopic objects are 
predictable I do not mean that the {\it existence} of such a matter of fact is 
predictable. Once again, a Born probability equal to~1 does not warrant the 
prediction {\it that} an event will happen or {\it that} a state of affairs will obtain. 
Only if it is {\it taken for granted} that exactly one of a range of possible events 
or states of affairs will happen or obtain, does a Born probability equal to~1 
allow us to predict {\it which} event or state of affairs will happen or obtain. 
What I mean by saying that matters of fact about the (successive) positions of 
a macroscopic object are predictable, is this: what an actual event or state of 
affairs implies regarding the position of a macroscopic object is consistent with 
what can be predicted with the help of some classical dynamical law on the 
basis of earlier position-defining events. Everything a macroscopic object does 
(that is, every matter of fact about its present properties) follows via the 
pertinent classical laws from what it did (that is, from matters of fact about its 
past properties).\fnote{%
The above definition of `macroscopic' does not stipulate that events indicating 
departures from the classically predicted behavior occur with zero {\it 
probability}. An object is entitled to the label `macroscopic' if no such event {\it 
actually} occurs during its lifetime. What matters is not whether such an event 
{\it may} occur (with whatever probability) but whether it ever {\it does} occur.}

When I speak of the {\it existence} of a matter of fact, I mean the occurrence of 
an actual event or the existence of an actual state of affairs. It is worth 
emphasizing that this is something that cannot be undone or `erased' (Englert, 
Scully, and Walther, 1999; Mohrhoff, 1999). According to Wheeler's 
interpretation of the Copenhagen interpretation, `no elementary quantum 
phenomenon is a phenomenon until it is registered, recorded, ``brought to a 
close'' by an ``irreversible act of amplification,'' such as the blackening of a 
grain of photographic emulsion or the triggering of a counter' (Wheeler, 1983). 
In point of fact, there is no such thing as an `irreversible act of amplification.' 
As long as what is `amplified' is counterfactuals, the `act of amplification' is 
reversible. No amount of amplification succeeds in turning a counterfactual into 
a fact. No matter how many counterfactuals get entangled, they remain 
counterfactuals. On the other hand, once a matter of fact exists, it is {\it 
logically} impossible to erase it. For the relevant matter of fact is not that the 
needle deflects to the left (in which case one could `erase' it by returning the 
needle to the neutral position). The relevant matter of fact is that {\it at a time} 
$t$ the needle deflects (or points) to the left. This is a timeless truth. If at the 
time $t$ the needle deflects to the left, then it always has been and always will 
be true that at the time $t$ the needle deflects to the left.

Note that an apparatus pointer is not a macroscopic object according to the 
above definition. In general there is nothing that allows one to predict which 
way the needle will deflect (given that it will deflect). Only {\it before} and {\it 
after} the deflection event does the needle behave as a macroscopic object. Is 
not such a definition self-defeating? It would be so if it were designed to explain 
why the needle deflects left {\it or} right (rather than both left {\it and} right). But 
such an explanation is neither required nor possible. If past events allow us to 
infer a superposition of the form 
$a\ket{\hbox{left}}\otimes\ket{a}+b\ket{\hbox{right}}\otimes\ket{b}$, they allow 
us to infer the following: {\it if} there is a matter of fact about the direction in 
which the needle deflects, it warrants the inference `left' with probability 
$\absosq{a}$, and it warrants the inference `right' with probability $\absosq{b}$. 
Nothing allows us to predict the existence of such a matter of fact. The 
deflection event is a causal primary, notwithstanding that it happens with a 
measurable probability, and that by a suitable choice of apparatus this 
probability can be made reasonably large.

As I have stressed elsewhere (Mohrhoff, 1999), what is true of particles in 
double-slit experiments is equally true of cats in double-door experiments. 
Except for the myriads of matters of fact about the door taken by the cat, `the 
door taken by the cat' is objectively undefined. This seems to entail a vicious 
regress. We infer the positions of particles from the positions of the detectors 
that click. But the positions of detectors are extrinsic, too. They are what they 
are because of the matters of fact from which one can (in principle) infer what 
they are. Thus there are detector detectors from which the positions of particle 
detectors are inferred, and then there are detectors from which the positions of 
detector detectors are inferred, and so on {\it ad infinitum}. However, as we 
regress from particle detectors to detector detectors and so on, we sooner or 
later (sooner rather than later) encounter a macroscopic detector whose 
position is not manifestly fuzzy. There the buck stops. The positions of things 
are defined in terms of the not manifestly fuzzy positions of macroscopic 
objects.

It is therefore consistent to think of the deflection of the pointer needle as one 
of those uncaused actual events on which the (contingent) properties of things 
depend. {\it Prima facie} we have another vicious regress: Like all contingent 
properties, the initial and final positions of the needle are what they are 
because of what happens or is the case in the rest of the world. They thus 
presupposes other `deflection events,' which presuppose yet other `deflection 
events,' and so on {\it ad infinitum}. But since before and after its deflection the 
needle behaves as a macroscopic object, its initial and final positions are 
quantitatively defined independently of what happens elsewhere. They are 
positions of the kind that are used to define positions. Hence the deflection 
event -- the transition from the initial to the final position -- is also independent 
of what happens elsewhere.

\section{\conclu}Language and the Indefinite

My chief conclusion in this paper is that (ER1) and (ER2) are both false. The 
sufficient and necessary condition for the existence of an element of reality 
$A=a$ is the existence of an actual state of affairs, or the occurrence of an 
actual event, from which $A=a$ can be inferred. The contingent properties of 
all quantum systems -- and in clear this means the positions of all material 
objects and whatever other properties can be inferred from them -- are 
extrinsic. They are defined in terms of the goings-on in the `rest of the world.' 
The reason why this does not send us chasing the ultimate property-defining 
facts in neverending circles, is the existence of a special class of objects the 
positions of which are not manifestly indefinite. Everything a macroscopic 
object does (that is, every matter of fact about its present properties) follows 
via the pertinent classical laws from what it did (that is, from matters of fact 
about its past properties). This makes it possible to ignore the fact that the 
properties of a macroscopic object, like all contingent properties, `dangle' from 
external events and/or states of affairs. Instead of having to conceive of the 
successive states of a macroscopic object as a bundle of statistically correlated 
inferences warranted by a multitude of causal primaries external to the object, 
we are free to think of the object's successive states as an evolving collection 
of intrinsic properties fastened only to each other, by causal links.

The familiar macroworld with its causal links and deterministic processes is 
something we project onto the fundamental statistical correlations and their 
uncaused correlata. This projection works where the correlations are not 
manifestly probabilistic (that is, where the statistical correlations evince no 
statistical variations). Diachronic correlations that are not manifestly 
probabilistic can be passed off as causal links. We can impose on them our 
agent-causality with some measure of consistency, even though this results in 
the application of a wrong criterion: temporal precedence takes the place of 
causal independence as the criterion which distinguishes a cause from its 
effect.

Quantum mechanics presupposes the macroworld: it assigns probabilities to 
conditionals that refer to events or states of affairs either in the actual 
macroworld or in a possible macroworld. This is the reason why Bohr (1934, 
1958) insisted not only on the necessity of describing quantum phenomena in 
terms of the experimental arrangements in which they are displayed, but also 
on the necessity of employing classical language in describing these 
experimental arrangements. Classical language is the language of causal 
processes, of definite states that evolve deterministically, of definite objects 
and of definite events -- in short, the language of the macroworld. Thus in one 
sense the microworld is fundamental (macroscopic objects are made of 
particles and atoms), and in another sense the macroworld is fundamental (the 
contingent properties of particles and atoms are defined in terms of the
goings-on in the macroworld). The mutual dependence of the quantum and 
classical `domains' has often been remarked upon (e.g., Landau and Lifshitz, 
1977), but I'm not sure it has been adequately appreciated.

It seems to me that what is ultimately responsible for this mutual dependence is 
the conflict between a real, objective indefiniteness and the intrinsic 
definiteness of language. Language is inherently `classical.' Discourse is of 
things -- the discrete carriers of significance that appear as the subjects of 
predicative sentences. Things fall into mutually disjoint classes according to the 
properties they possess or lack. For any two different classes $C_1$ and 
$C_2$ there exists a property $p$ such that `$x$ has $p$' is true of all 
members of $C_1$ and `$x$ lacks $p$' is true of all members of $C_2$. This 
seems to warrant the following Principle of Completeness (Wolterstorff, 1980): 
for every thing $x$ and every property $p$, $x$ either has $p$ or lacks $p$. 
Reality, however, doesn't play along with this linguistic requirement. 
Sometimes `$x$ has $p$' is neither true nor false but meaningless. There are 
situations in which nothing in the real world corresponds to the linguistic (or 
conceptual) distinction between `$x$ has $p$' and `$x$ lacks $p$'. In such 
situations it is nevertheless meaningful to consider what would have happened 
if one had found out whether $x$ has $p$ or lacks $p$, and to assign objective 
probabilities to the alternatives `$x$ has $p$' and `$x$ lacks $p$.'

Given the intrinsic definiteness of language, the natural way to express an 
objective indefiniteness is to use counterfactuals. One then has one 
counterfactual for each alternative (`if $Q$ were measured, the result would be 
$q_k$'), and at least one of them comes with a nontrivial objective probability 
(that is, an objective probability other than 0 or 1). The linguistic requirement of 
definiteness is met by the use of counterfactuals the respective consequents of 
which conform to the Principle of Completeness: each consequent explicitly 
affirms the truth of one alternative and implicitly denies the truth of the other 
alternatives. The objective indefiniteness finds expression in the fact that the 
counterfactuals are assigned nontrivial objective probabilities rather than truth 
values.

Objective indefiniteness thus leads to the use of counterfactuals with nontrivial 
objective probabilities, and nontrivial objective probabilities, as we have seen, 
entail that the properties affirmed by the counterfactuals' consequents are 
extrinsic: they are defined in terms of the goings-on in the macroworld, 
notwithstanding that the objects of the macroworld are made up -- or shall we 
say, manifested by means -- of nonmacroscopic objects. This mutual 
dependence of the two `domains' would amount to a vicious circle if the 
properties of the macroworld were in their turn defined in terms of the 
microworld. But this is not the case.

Since the contingent properties of things are defined in terms 
of events or states of affairs in the macroworld, quantum mechanics 
presupposes the macroworld. In particular, it presupposes such matters of fact as `the needle deflects to the left' or `the needle is pointing left.' By itself this does not guarantee
that quantum mechanics is consistent with the existence of the macroworld - quantum mechanics 
(or the interpretation of quantum mechanics put forward in this paper) could lack
self-consistency. But self-consistency only requires that the needle's 
position too is fuzzy, and that it `dangles ontologically' from the goings-on in the 
rest of the world. Quantum mechanics permits it to `dangle' from them in such a way that, before and after the deflection, it is not manifestly fuzzy. If the needle's position is not manifestly fuzzy, the needle behaves 
as a macroscopic object, and we can consistently conceive of its successive 
positions as `dangling causally' from each other -- except for one gap in the 
causal chain, the deflection event. But being a (probabilistic) transition between 
states embedded in the causal nexus of the macroworld, this too forms part of 
the macroworld.

Quantum mechanics not only presupposes and admits of the existence of macroscopic objects, it also entails it. The existence of an unpredictable matter of fact about the position 
of $O$ entails the existence of detectors with `sharper' positions; the existence 
of an unpredictable matter of fact about the position of one of those detectors 
entails the existence of detectors with yet `sharper' positions; and so on. It 
stands to reason that one sooner or later runs out of detectors with `sharper' 
positions. There are `ultimate' detectors the positions of which are not 
manifestly fuzzy, and which therefore are macroscopic.

\vfill\break

\parindent=0pt\bigskip{\bf References}\par\smallskip
\baselineskip=14truept\parskip=3truept

Aharonov, Y., Bergmann, P.G., and Lebowitz, J.L. (1964) `Time Symmetry in 
the Quantum Process of Measurement,' {\it Physical Review B} {\bf 134},
1410-1416.

Aharonov,Y., and Vaidman, L. (1991) `Complete Description of a Quantum 
System at a Given Time,' {\it Journal of Physics} {\bf A 24}, 2315-2328.

Albert, D.Z. (1992) {\it Quantum Mechanics and Experience}, Cambridge, MA: 
Harvard University Press.

Bartley III., W.W. (1982) {\it Quantum Theory and the Schism in Physics}, 
Totowa, NJ: Rowan \& Littlefield.

Bell, J.S. (1987) {\it Speakable and Unspeakable in Quantum Mechanics}, 
Cambridge: Cambridge University Press.

Bell, J.S. and Nauenberg, M. (1966) `The Moral Aspect of Quantum 
Mechanics,' in De Shalit, Feshbach, and Van Hove (1996), pp. 279-86. 
Reprinted in Bell (1987), pp. 22-28.

Bohm, D. (1951) {\it Quantum Theory}, Englewood Cliffs, NJ: Prentice 
Hall.

Bohr, N. (1934) {\it Atomic Theory and the Description of Nature}, Cambridge: 
Cambridge University Press.

Bohr, N. (1958) {\it Atomic Physics and Human Knowledge}, New York: Wiley, 
p.~72.

Bohr, N. (1963) {\it Essays 1958-62 on Atomic Physics and Human 
Knowledge}, New York: Wiley, p.~3.

Cassinello, A., and S\'anchez-G\'omez, J.L. (1996) `On the Probabilistic 
Postulate of Quantum Mechanics,' {\it Foundations of Physics} {\bf 26}, 
1357-1374.

Davies, P. (1989) {\it The New Physics}, Cambridge: Cambridge University 
Press.

De Shalit, A., Feshbach, H., and Van Hove, L. (1966) {\it Preludes in 
Theoretical Physics}, Amsterdam: North Holland.

d'Espagnat, B. (1976) {\it Conceptual Foundations of Quantum Mechanics}, 2nd 
edition, Reading, MA: Benjamin, p. 251.

d'Espagnat, B. (1995) {\it Veiled Reality}, Reading, MA: Addison-Wesley.

Einstein, A., Podolsky, B., and Rosen, N. (1935) `Can Quantum-Mechanical 
Description of Physical Reality be Considered Complete?,'' {\it Physical 
Review} {\bf 47}, 777-780.

Englert, B.-G., Scully, M.O., and Walther, H. (1994) `The Duality in Matter and 
Light,' {\it Scientific American} {\bf 271}, No. 6 (December), 56-61.

Englert, B.-G., Scully, M.O., and Walther, H. (1999) `Quantum Erasure 
in Double-Slit Interferometers with Which-Way Detectors,'' {\it American 
Journal of Physics} {\bf 67}, 325-329.

Ford, J., and Mantica, G. (1992) `Does Quantum Mechanics Obey the 
Correspondence Principle? Is it Complete?', {\it American Journal of Physics} 
{\bf 60}, 1068-1097.

Gell-Mann, M., and Hartle, J.B. (1990) `Quantum Mechanics in the Light of 
Quantum Cosmology,' in Zurek (1990), pp. 425-458.

Ghirardi, G.C., Rimini, A., and Weber, T. (1986) `Unified Dynamics for 
Microscopic and Macroscopic Systems,' {\it Physical Review D} {\bf 34}, 
470-91.

Gisin, N., and Percival, I.C. (1992) `The Quantum-State Diffusion Model 
Applied to Open Systems,' {\it Journal of Physiccs A} {\bf 25}, 5677-5691.

Griffiths, R.B. (1984) `Consistent Histories and the Interpretation of Quantum 
Mechanics,' {\it Journal of Statistical Physics} {\bf 36}, 219-272.

Heisenberg, W. (1958) {\it Physics and Philosophy}, New York: Harper and 
Row, Chapter 3.

Hilgevoord, J. (1998) `The Uncertainty Principle For Energy and Time. II,' {\it 
American Journal of Physics } {\bf 66}, 396-402.

Jauch, J.M. (1968) {\it Foundations of Quantum Mechanics}, Reading, 
MA: Addison-Wesley.

Joos, E., and Zeh, H.D. (1985) `The Emergence of Classical Properties 
Through Interaction With the Environment,' {\it Zeitschrift f\"ur Physik B} {\bf 
59}, 223-243.

Kastner, R.E. (1999) `Time-Symmetrized Quantum Theory, Counterfactuals, 
and ``Advanced Action'',' to be published in {\it Studies in History and 
Philosophy of Science}.

Landau, L.D., and Lifshitz, E.M. (1977) {\it Quantum Mechanics}, Oxford: 
Pergamon Press, pp.~2-3.

Langevin, P. (1939) {Actualit\'es scientifiques et industrielles: Expos\'es de 
physique g\'en\'erale}, Paris: Hermann.

Lockwood, M. (1989) {\it Mind, Brain and the Quantum: The Compound `I'}, 
Oxford: Basil Blackwell.

London, F., and Bauer, E. (1939) `La th\'eorie de l'observation en m\'ecanique 
quantique,' in Langevin (1939), No. 775. English translation `The Theory of 
Observation in Quantum Mechanics' in Wheeler and Zurek (1983), pp.
217-259.

L\"uders, G. (1951) `\"Uber die Zustands\"anderung durch den 
Messprozess,' {\it Annalen der Physik (Leipzig)} {\bf 8}, 322-328.

Mermin, N.D. (1998) ``What is Quantum Mechanics Trying to Tell Us?,'' {\it 
American Journal of Physics} {\bf 66}, 753-767.

Miller, A.I. (1990) {\it 62 Years of Uncertainty}, New York: Plenum Press.

Mohrhoff, U. (1999) `Objectivity, Retrocausation, and the Experiment of 
Englert, Scully and Walther,' {\it American Journal of Physics} {\bf 67},
330-335.

Mohrhoff, U. (submitted) `What Quantum Mechanics is Trying to Tell Us.'

Omn\`es, R. (1992) `Consistent Interpretations of Quantum Mechanics,' {\it 
Reviews of Modern Physics} {\bf 64}, 339-382.

Onley, D., and Kumar, A. (1992) `Time Dependence in Quantum 
Mechanics -- Study of a Simple Decaying System,' {\it American Journal 
of Physics} {\bf 60}, 432-439.

Page, D.N. (1996) `Sensible Quantum Mechanics: Are Probabilities Only in the 
Mind?,' {\it International Journal of Modern Physics D} {\bf 5}, 583-596.

Page, D.N., and Wootters, W.K. (1983) `Evolution Without Evolution: 
Dynamics Described by Stationary Observables' {\it Physical Review D} {\bf 
27}, 2885-2891.

Pais, A., (1982) {\it `Subtle is the Lord...': The Science and the Life of Albert 
Einstein}, Oxford: Clarendon Press.

Pearle, P. (1989) `Combining Stochastic Dynamical State-Vector Reduction 
with Spontaneous Localization,' {\it Physical Review A} {\bf 39}, 2277-2289.

Percival, I.C. (1994) `Primary State Diffusion,' {\it Proceedings of the Royal 
Society London} {\bf A447}, 189-209.

Peres, A., and Zurek, W.H. (1982) `Is Quantum Theory Universally Valid?,' {\it 
American Journal of Physics} {\bf 50}, 807-810.

Popper, K.R. (1982) in Bartley III (1982).

Price, H. (1996) {\it Time's Arrow \& Archimedes' Point}, New York: Oxford 
University Press.

Primas, H. (1990) `Mathematical and Philosophical Questions in the Theory of 
Open and Macroscopic Quantum Systems,' in Miller (1990), pp. 233-257.

Redhead, M. (1987) {\it Incompleteness, Nonlocality and Realism}, Oxford: 
Clarendon, p. 72.

Shimony, A. (1978) `Metaphysical Problems in the Foundations of Quantum 
Mechanics,' {\it International Philosophical Quarterly} {\bf 18}, 3-17.

Shimony, A. (1989) `Conceptual Foundations of Quantum Mechanics' in Davies 
(1989), 373-95.

Scully, M.O., Englert, B.-G., and Walther, H. (1991) `Quantum Optical Tests of 
Complementarity,' {\it Nature} {\bf 351}, No. 6322, 111-116.

Singh, I., and Whitaker, M.A.B. (1982) `Role of the Observer in Quantum 
Mechanics and the Zeno Paradox,' {\it American Journal of Physics} {\bf 
50}, 882-887.

Stapp, H.P. (1993) {\it Mind, Matter, and Quantum Mechanics}, Berlin: 
Springer.

Vaidman, L. (1993) `Lorentz-Invariant ``Elements of Reality'' and the Joint 
Measurability of Commuting Observables,' {\it Physical Review Letters} {\bf 
70}, 3369-3372.

Vaidman, L. (1996a) `Defending Time-Symmetrized Quantum Theory,' e-print 
archive quant-ph 9609007.

Vaidman, L. (1996b) `Weak-Measurement Elements of Reality,' {\it 
Foundations of Physics} {\bf 26}, 895-906.

Vaidman, L. (1997) `Time-Symmetrized Counterfactuals in Quantum Theory,' 
e-print archive quant-ph 9807075, Tel-Aviv University Preprint TAUP-2459-97.

Vaidman, L. (1998) `Time-Symmetrized Quantum Theory,' {\it Fortschritte der 
Physik} {\bf 46}, 729-739.

Vaidman, L. (1999) `Defending Time-Symmetrized Quantum Counterfactuals,' 
to be published in {\it Studies in History and Philosophy of Science}.

von Neumann, J. (1955) {\it Mathematical Foundations of Quantum 
Mechanics}, Princeton: Princeton University Press.

Wheeler, J.A. (1983) `On Recognizing ``Law Without Law'' (Oersted 
Medal Response at the Joint APS-AAPT Meeting, New York, 25 January 
1983),' {\it American Journal of Physics} {\bf 51}, 398-404.

Wheeler, J.A., and Zurek, W.H. (1983) {\it Quantum Theory and 
Measurement}, Princeton, NJ: Princeton University Press.

Wolterstorff, N. (1980), {\it Works and Worlds of Art}, Oxford: Clarendon Press.

Zurek, W.H., (1981) `Pointer Basis of Quantum Apparatus: Into What Mixture 
Does the Wave Packet Collapse?', {\it Physical Review D} {\bf 24}, 1516-1525.

Zurek, W.H., (1982) `Environment-Induced Superselection Rules,' {\it Physical 
Review D} {\bf 26}, 1862-1880.

Zurek, W.H., (1990) {\it Complexity, Entropy, and the Physics of Information}, 
Reading, MA: Addison-Wesley.

\bye